\magnification=1000
\headline{\hfill\folio}
\hsize=15.0 true cm
\vsize=22.0 true cm
\baselineskip=13 pt
\input psfig.sty 
\def\ref{\par\noindent\hangindent=1.0 true cm}
\nopagenumbers
\centerline{\bf HELIUM CONTAMINATION FROM THE PROGENITOR STARS}
\centerline{\bf OF PLANETARY NEBULAE: THE He/H RADIAL GRADIENT AND}
\centerline{\bf THE $\Delta Y/\Delta Z$ ENRICHMENT RATIO
    \footnote \dag {\sevenrm Accepted for publication in Astrophysics and Space Science}}
\bigskip
\bigskip
\centerline{WALTER J. MACIEL}
\bigskip
\centerline{IAG/USP, S\~ao Paulo, Brazil}
\bigskip
\centerline{\it maciel@iagusp.usp.br}
\centerline{\it http://www.iagusp.usp.br/$\sim$maciel}
\bigskip
\bigskip
\noindent 
\centerline{\bf ABSTRACT}
\bigskip
\vbox{\baselineskip 11 pt\noindent
In this work, two aspects of the chemical 
evolution of $\, ^4$He in the Galaxy are considered on the basis 
of a sample of disk planetary nebulae (PN). First, an application 
of corrections owing to the contamination of \ $^4$He from the 
evolution of the progenitor stars shows that the He/H abundance 
by number of atoms is reduced by $0.012$ to $0.015$ in average, 
leading to an essentially flat He/H radial 
distribution. Second, a determination of the helium to heavy element 
enrichment ratio using the same corrections leads to values in
the range $2.8 < \Delta Y/\Delta Z < 3.6$ for $Y_p = 0.23$ and 
$2.0 < \Delta Y/\Delta Z < 2.8$ for $Y_p = 0.24$, in good agreement 
with recent independent determinations and theoretical models.}
\bigskip
\bigskip
\centerline{\bf 1. Introduction}
\medskip\noindent
The chemical evolution of \ $^4$He in the Galaxy can be studied
on the basis of abundance determinations in photoionized nebulae,
namely HII regions and planetary nebulae (Peimbert, 1990, 1995; 
Pagel, 1989, 1995). Also, recent 
work on HII regions in blue compact galaxies has given many 
informations on this problem, especially regarding the pregalactic 
helium abundance $Y_p$ (Pagel 1995, Peimbert 1995). 
Recent determinations of this quantity 
are usually in the range $Y_p = 0.23$ to $0.24$, and in fact most 
discussions now concentrate on the third decimal place (Olive 
{\it et al.} 2000, Burles {\it et al.} 1999, 
Izotov {\it et al.} 1999). On the other hand, some 
aspects of the chemical evolution of \ $^4$He have much greater 
uncertainties, particularly the question of the He/H radial gradient, 
$d({\rm He/H})/dR$, and the helium to metals enrichment ratio, 
$\Delta Y/\Delta Z$.

Planetary nebulae (PN) are also interesting in the study of these 
problems, since their He/H abundances are relatively well measured, 
with estimated uncertainties better than 20\%,  and large samples are 
available, spanning a large metallicity range. However, as the 
offspring of intermediate mass stars, these nebulae can have chemical 
compositions that are different from the original interstellar 
abundances. Therefore, in order to study the evolution of elements 
such as $\, ^4$He on the basis of these objects, it is necessary to 
take into account the He contamination by the PN progenitor stars. 
Previous work on these questions has introduced this contamination 
in an approximate way (Chiappini and Maciel 1994 or
not at all (Maciel 1988, Maciel and Chiappini 1994). 
In the past few years, however, detailed 
determinations have been published of the He yields and the amount
of this element produced during the dredge-up episodes that occur 
during the late evolution of intermediate mass stars, both as a 
function of the stellar mass and metallicity. 

In the present work, we make use of these calculations in order to 
estimate the $\, ^4$He contamination for a sample of disk planetary 
nebulae. Taking into account some relationships involving the 
nebular abundances, the central star mass and the stellar mass on 
the main sequence, such contamination can be {\it individually} 
determined. As a consequence, the analysis of the He/H radial 
gradient and of the enrichment ratio can be performed on more solid 
grounds. In section~2 the question of the abundance gradients is 
considered, and in section~3 the $\Delta Y/\Delta Z$ ratio is discussed.
In section~4 a simple model to account for the contamination of the 
progenitor stars is proposed, and the results are presented and 
discussed in section~5.
\bigskip
%
\centerline{\bf 2. Radial abundance gradients}
\medskip\noindent 
Radial abundance gradients of element ratios such as O/H and
S/H are well known, and amount to $d\log ({\rm O/H})/dR
\simeq -0.07$ dex/kpc (Maciel, 1997, 2000). 
These gradients can be derived from a variety of objects, comprising 
HII regions, planetary nebulae, B stars, etc. The study of these
gradients includes a determination of their average magnitude,
the existence of spatial variations along the disk and the
time variation of the gradients, as the objects mentioned have 
different ages, spanning a few Gyr. Several mechanisms have been 
proposed to account for the gradients, and the reader is referred
to Matteucci and Chiappini (1999) for a recent review 
on this topic.

The existence of a He/H gradient is much more uncertain. Shaver
et al. (1983) obtained a zero gradient for the ratio 
He$^+$/H$^+$ from radio and optical measurements in galactic HII regions,
and suggested a small gradient, $d\log({\rm He/H})/dR = -0.001$
dex/kpc. Their derived dispersion was $\sigma_d \simeq 0.08$, higher 
than the estimated observational untertainty, $\sigma_{obs}
\simeq 0.01$ to $0.02$. Earlier work suggested a negative
gradient, $d({\rm He/H})/dR \simeq -0.005$ to $-0.008$ kpc$^{-1}$, 
corresponding approximately to $d\log({\rm He/H})/dR \simeq -0.02$ 
to $-0.03$ dex/kpc (D'Odorico {\it et al.} 1976,
Peimbert and Serrano 1980). 

Fa\'undez-Abans and Maciel (1986) obtained an average
gradient of $d({\rm He/H})/dR  \simeq -0.005$ kpc$^{-1}$ or 
$d\log({\rm He/H})/dR \simeq -0.02$ dex/kpc for a sample 
of galactic PN. Pasquali and Perinotto (1993) 
obtained negligible gradients, $d({\rm He/H})/dR \simeq -0.001$ 
kpc$^{-1}$ for a sample of disk PN, reaching a maximum value 
$d({\rm He/H})/dR \simeq -0.004$~kpc$^{-1}$ only if objects 
close to the galactic disk ($z \simeq 0$) were included, but in 
this case the galactocentric distance variation was small and 
the result was uncertain. Considering PN of types I and II 
(Peimbert 1978) with distances from the galactic
plane $z < 300$~pc, they obtained a gradient $d({\rm He/H})/dR 
\simeq -0.003$ kpc$^{-1}$, corresponding approximately to 
$d(\log{\rm He/H})/dR \simeq -0.009$ dex/kpc. Amnuel 
(1993) obtained gradients generally in the range 
$d\log({\rm He/H})/dR \simeq -0.002$ to $-0.026$ dex/kpc for the 
different classes of his adopted classification system. A discussion 
of these results has been given by Esteban and Peimbert 
(1995).  Maciel and Chiappini (1994) 
also obtained very small gradients, $d({\rm He/H})/dR \simeq -0.0004 
\pm 0.0006$ kpc$^{-1}$ for a sample of type II PN with known distances. 
Recently, Esteban {\it et al.} (1999) derived a very 
small gradient, $d\log({\rm He/H})/dR \simeq -0.004 \pm 0.005$ dex/kpc, 
corresponding to $d({\rm He/H})/dR \simeq -0.001$ kpc$^{-1}$ based on 
He recombination lines of three HII regions (M8, M17 and Orion). 
This result is apparently independent of temperature 
fluctuations in the nebulae.

All these determinations involving PN until now have {\it not} taken 
into account the He contamination by the progenitor stars during 
their evolution. It is well known that main sequence stars in
the mass interval $1 \leq M/M_\odot \leq 8$ have an
important contribution to the He production, especially
during the first and second dredge-up episodes (cf. Peimbert
1990). Therefore, a realistic determination of the He/H 
gradient from planetary nebulae should take into account the He 
production of their progenitor stars. 
\bigskip
\bigskip
\centerline{\bf 3. The helium to metals enrichment ratio}
\medskip\noindent
The helium to metals enrichment ratio is an important parameter
in the study of the chemical evolution of the Galaxy, and  is
usually determined from photoionized nebulae such as HII regions
and HII galaxies (Lequeux {\it et al.} 1979, Peimbert 
1990, 1995, Pagel {\it et al.} 1992, Izotov {\it et al.} 1997, Thuan and 
Izotov 1998, Esteban {\it et al.} 1999). 
Recent work has also taken into account the fine structure in 
the main sequence of nearby stars (Pagel and Portinari 
1998, H\o g {\it et al.} 1998) and 
Hipparcos parallaxes for open clusters have been used to stress 
local variations of this ratio (Efremov {\it et al.} 1997). 
Results obtained from these methods are in the range 
$2 \leq \Delta Y/\Delta Z \leq 6$, and lower values are generally 
in better agreement with theoretical calculations.

Planetary nebulae can also be used to determine the 
$\Delta Y/\Delta Z$ ratio, and earlier determinations of this 
quantity include D'Odorico {\it et al.} (1976) and 
Peimbert and Serrano (1980), with results in the range 
$2.2 < \Delta Y/\Delta Z < 3.6$.

Maciel (1988) determined the pregalactic abundance and the
helium to metals enrichment ratio using a sample of type~II
PN, assuming that the He contamination from the central star
was negligible, with the result $\Delta Y/\Delta Z \simeq 3.5 \pm 0.3$. 
Chiappini and Maciel (1994) made a first attempt to take into 
account the contamination from the PN progenitor stars. For these
objects, the observed He abundances by mass $Y$ can be written as

$$Y = Y_p + {\Delta Y \over \Delta Z} \ Z + \Delta Y_s \eqno(1)$$

\noindent 
(Peimbert and Torres-Peimbert, 1974, 1976), 
where $Y_p$ is the pregalactic abundance, $Z$ is the heavy element 
abundance by mass and $\Delta Y_s$ is the contamination from the 
progenitor star. In the work of Chiappini and Maciel (1994), 
the  contamination $\Delta Y_s$ was included as a free parameter based 
on calculations by Boothroyd (private communication). The adopted 
values were $\Delta Y_s = 0.0, 0.008, 0.015$ and $0.022$, which were 
applied to all nebulae in the sample. From direct $Y(Z)$ correlations, 
the derived ratio was in the range $3.4 \leq \Delta Y/\Delta Z \leq 5.6$ 
[see Table~3 of Chiappini and Maciel (1994)]. The results by
Chiappini and Maciel (1994) were based on {\it average} 
corrections, applied to all nebulae, so that any differences in the 
individual behaviour of the nebulae were lost. Since there is now some 
information on the masses of the PN central stars and the corresponding
masses on the main sequence, it is interesting to revise these
results in order to account for  the contamination of the 
progenitor stars {\it individually}.
\bigskip
\bigskip
\centerline{\bf 4. He production in intermediate mass stars}
\medskip\noindent
Since the pioneering work of Renzini and Voli (1981), several
investigators have presented detailed calculations of yields of the
different elements in intermediate mass stars, based on recent
evolutionary models (see for example Maeder 1992, van den 
Hoek and Groenewegen 1997, Boothroyd and Sackmann 1999, 
Marigo 2000). Regarding He abundances, these models take 
into account the dredge-up episodes that occur in intermediate mass stars,
particularly the first and second processes. In this work, we
have adopted the recent yields by van den Hoek and Groenewegen 
(1997) for stars with masses in the range $0.9$ to  $7\, M_\odot$ 
on the main sequence. Their results include all dredge up episodes
and are based on evolutionary tracks of the Geneva group up to the 
AGB branch, coupled with a synthetical AGB model with hot bottom 
burning. The yields depend on the stellar mass and metallicity, and 
the standard model takes as parameters the mass loss scaling parameter, 
$\eta_{AGB} = 4$, the minimum core mass for dredge up, $M_c^{min} = 0.58$ 
and the third dredge up efficiency, $\lambda = 0.75$ (see van den Hoek 
and Groenewegen 1997 for a detailed discussion).

As a comparison with the van den Hoek and Groenewegen (1997) data,
we have used the computations by Boothroyd (private communication, see 
also Boothroyd and Sackmann 1999, Sackmann and Booth\-royd 
1999)  based on earlier theoretical models by Boothroyd and 
Sackmann (1988). These computations are restricted to the first
and second dredge up processes, and show some decrease in the He 
contamination for masses roughly between one and three solar masses.
As a consequence, they produce smaller corrections for stellar 
masses under $4\, M_\odot$ than the previous set, so that they
are used for comparison purposes. 

The He excess by mass $\Delta Y_s$ can be estimated from the yields
of the intermediate mass stars as a function of the stellar mass 
on the main sequence $M_{MS}$ and total metallicity $Z$. We have 
considered here $Z = 0.020$, which is appropriate for type II PN in 
the galactic disk, but different metallicities do not significantly 
change our results. In this work, we made the simplifying assumption
that most of the He excess produced by the progenitor stars and dredged
up to the surface is mixed up in the outer layers and ejected into 
the nebulae. Therefore, the determination of the corrections is
straightforward, and our derived contamination $\Delta Y_s$  
corresponds to the largest possible correction for a given mass.

In order to obtain the main sequence mass of the PN progenitor
stars, we have adopted a two step procedure. First, we obtained
the central star mass $M_c$ from the observed N/O nebular
abundance, and then we have used an initial mass-final mass
relation for intermediate mass stars in order to derive $M_{MS}$.

The existence of a relationship between the PN core mass and the
N/O abundance has been discussed by a number of people, often with
conflicting results (see Cazetta and Maciel 2000 for some
references). Until recently, theoretical models were unable to predict
some measurable increase in the N/O and He/H ratios for PN
with relatively massive progenitors, a situation that has
changed with the extension of the models up to 7 solar masses
on the main sequence with overshooting (Marigo 2000) 
or by the inclusion of hot bottom burning (van den Hoek and Groenewegen
1997). A calibrated relation has been recently proposed by
Cazetta and Maciel (2000) as

$$M_c = a + b \ \log({\rm N/O}) + c \ [\log({\rm N/O})]^2 \eqno(2)$$

\noindent
where $M_c$ is in solar masses and N/O is the abundance of nitrogen
relative to oxygen by number of atoms. We have adopted $a = 0.689$, 
$b = 0.056$ and $c = 0.036$ for $-1.2 \leq \log ({\rm N/O}) \leq -0.26$ 
and $a = 0.825$, $b = 0.936$ and $c = 1.439$ for 
$\log ({\rm N/O}) > -0.26$. This relation is valid in the approximate 
interval $-1 < \log{\rm (N/O)} < 0.2$, and has been calibrated using 
synthetic models for AGB stars by Groenewegen and de Jong 
(1993) and Groenewegen {\it et al.} (1995), 
and NLTE model atmospheres by M\'endez et al. (1988, 1992). 
Since the derived core masses are relatively 
large, $M_c \geq 0.67 M_\odot$, leading to relatively large main sequence 
masses, we will refer to this calibration as {\it high-mass calibration}.

Alternatively, we have adopted a different calibration which
leads to core masses and main sequence masses lower than the values 
quoted above. In fact, recent work on the mass distribution of PN 
central stars (Stasi\'nska {\it et al.} 1997, 
Stasi\'nska and Tylenda 1990) based on self-consistent 
and distance independent methods, suggest that more than 80\% of the PN 
central stars have masses in the range $M_c \simeq 0.55$ to $0.65\,M_\odot$. 
These results are also in agreement with core masses derived by Zhang 
(1993) and with the mass-N/O abundance relation by
Marigo et al. (1996). This calibration can be written as

$$M_c = 0.7242 + 0.1742 \ \log ({\rm N/O}) \eqno(3)$$

\noindent
and replaces Eq.~2 for $\log ({\rm N/O}) \leq -0.26$. Since 
this calibration produces lower core masses, $M_c \geq 0.55\, M_\odot$ 
and lower main sequence masses  than the first calibration, we will 
refer to it as {\it low-mass calibration}.

The average initial mass-final mass relation for the {\it high-mass
calibration} was taken from the gravity distance work of
Maciel and Cazetta (1997), and can be written as

$$M_c = a_0 + a_1 \ M_{MS} + a_2 \ M_{MS}^2 + a_3 \ M_{MS}^3
    + a_4 \ M_{MS}^4 \eqno(4)$$

\noindent
where $a_0 = 0.5426$, $a_1 = 0.02093$,  $a_2 = -0.01122$,
$a_3 = 0.00447$ and  $a_4 = -0.0003119$ (see figure~1 of Maciel
and Cazetta 1997). This relation was based on models by 
Sch\"onberner (1983), Bl\"ocker and Sch\"onberner 
(1990), Groenewegen and de Jong (1993) and 
Weidemann and Koester (1983), and can be used in 
the whole range of intermediate star masses, namely 
$0.8 < M_{MS}/M_\odot < 7$. For details the reader is
referred to Maciel and Cazetta (1997). Used in conjunction 
with Eq.~2, this equation favours relatively large main
sequence masses, $M_{MS} \geq 3\, M_\odot$, in agreement with the
NLTE model atmospheres of M\'endez {\it et al.} (1988, 1992).

For the {\it low-mass calibration}, we have adopted the following
initial mass-final mass relation

$$M_c = 0.4877 + 0.0623 \ M_{MS} \eqno(5)$$

\noindent
which favours lower masses, $M_{MS} \geq 1 M_\odot$. This relation 
is in better agreement with the corresponding relation of 
Groenewegen and de Jong (1993) for $M_{MS} < 3 M_\odot$ and 
with the relations by Sch\"onberner (1983) and Bl\"ocker and
Sch\"onberner (1990) for higher masses. It also produces
main sequence masses closer to the values originally 
proposed for type II PN by Peimbert (1978).
\bigskip
\bigskip
\centerline{\bf 5. Results and discussion}
\bigskip\noindent 
{5.1 THE He/H RADIAL GRADIENT}
\medskip\noindent
We have applied the procedure outlined in section~4 to a sample of 
disk planetary nebulae used by Maciel and Quireza (1999) 
to study radial abundance gradients of O/H, Ne/H, S/H and Ar/H. 
This sample includes basically the objects of Maciel and K\"oppen 
(1994), Maciel and Chiappini (1994), 
Costa {\it et al.} (1996, 1997) and a few 
objects not included in the previous samples. The set of abundances
adopted by Maciel and Quireza (1999) is not a 
homogeneous one, in the sense that they have been derived by
different groups using different computational procedures, such
as icf's or a different treatment of the collisional excitation
of He lines. However, as previously discussed by Maciel and Quireza 
(1999) and Maciel and K\"oppen (1994),
the final sample includes only the best determined abundances, 
for which at least two recent measurements are available, so that
the adopted abundances have uncertainties as low as possible.
A few objects are low excitation nebulae, which may
have some contribution of neutral He, but this paper concerns
basically with high excitation objects, which are located at least
at 4 kpc from the galactic centre. Furthermore, we have
preferentially selected objects studied by our own group,
to keep the inhomogeneity of the sample at a minimum level.
For details on the observational and reduction procedures 
by the IAG/USP group the reader is referred to Costa {\it et al.} 
(1996, 1997) and references therein. 

The initial sample contained 130 PN. We have removed 11 objects 
without reliable He/H or N/O abundances and 16 PN with very high N/O 
values, thus partially avoiding the effects of temperature fluctuations 
(cf. Peimbert 1995, Gruenwald and Viegas 1998) 
and also the large and uncertain corrections needed for these objects. 
In fact, Gruenwald and Viegas (1998) have presented
some empirical evidence that the higher stellar temperatures
characteristic of Type~I PN are associated with higher helium
and nitrogen abundances, increasing the effects of temperature
fluctuations. Distances and abundances with sources are given 
by Maciel and Quireza (1999). The final sample 
contains 103 PN and is shown in Tables~I  to III, 
which include (i) the common name of the nebula, (ii) the 
galactocentric distance (kpc) adopting $R_0 = 7.6$ kpc as in Maciel 
and Quireza (1999), see also Maciel (1993), 
(iii) to (v) the observed abundances of He, N and O, given as 
He/H, $\log$(N/O) and $\epsilon({\rm O}) = \log$(O/H)+12, 
and (vi) the references for the abundances, which are listed
after the tables. 

The uncorrected He/H abundances are shown in Fig.~1 
(top panel) as a function of the galactocentric distance $R$. The 
straight line shows a least squares linear fit. It can be seen that 
no gradient is present, and that the average abundance is 
He/H $\simeq 0.106 \pm 0.003$. 
\bigskip
%
\centerline{\psfig{figure=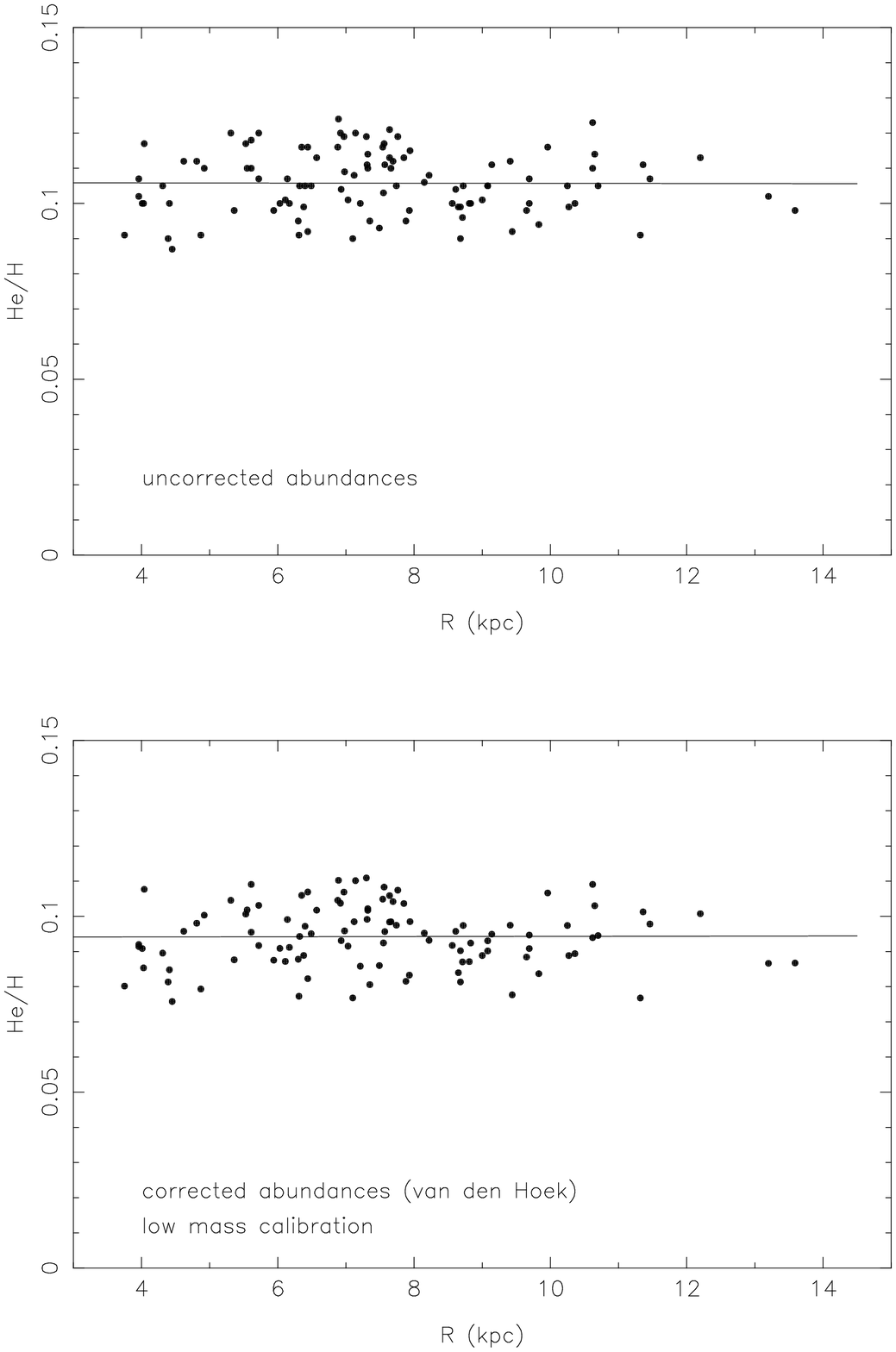,height=16.0 cm,angle=0} }
\noindent{\sevenrm \baselineskip 12 pt Figure 1 -  
    Top panel: uncorrected He/H 
    abundances as a function 
    of the galactocentric distance and least squares linear fit.
    Bottom panel: The same using corrections for the low mass 
    calibration.}
%
\vfill\eject

\hskip 3.0 true cm
Table I - Input data for PN
$$\vbox{\halign{#\hfil&\quad\hfil#\hfil&\quad\hfil#\hfil
&\quad\hfil#\hfil&\quad\hfil#\hfil&\quad\hfil#\hfil\cr\noalign{\hrule}
\ \cr
name & $R$ & He/H & $\log$(N/O) & $\epsilon$(O) & Ref. \cr
\ \cr
\noalign{\hrule}
\ \cr
NGC 1535    &  8.71 & 0.096 & -1.07 & 8.61 & 1   \cr
NGC 2022    &  9.69 & 0.107 & -0.53 & 8.41 & 1   \cr
NGC 2371    &  9.00 & 0.101 & -0.52 & 8.63 & 1   \cr
NGC 2392    &  9.44 & 0.092 & -0.12 & 8.50 & 1   \cr
NGC 2438    &  8.61 & 0.104 & -0.69 & 8.84 & 1   \cr
NGC 2452    &  9.14 & 0.111 & -0.15 & 8.61 & 6   \cr
NGC 2792    &  7.94 & 0.115 & -0.20 & 8.75 & 1   \cr
NGC 2867    &  7.54 & 0.116 & -0.60 & 8.71 & 1   \cr
NGC 3132    &  7.64 & 0.121 & -0.46 & 8.95 & 1,2 \cr
NGC 3195    &  6.89 & 0.124 & -0.52 & 8.90 & 1   \cr
NGC 3211    &  7.30 & 0.119 & -0.77 & 8.70 & 1   \cr
NGC 3242    &  7.74 & 0.105 & -0.75 & 8.66 & 1   \cr
NGC 3587    &  7.93 & 0.098 & -0.30 & 8.59 & 1   \cr
NGC 3918    &  6.98 & 0.109 & -0.50 & 8.78 & 1   \cr
NGC 5307    &  6.30 & 0.095 & -0.74 & 8.64 & 1   \cr
NGC 5882    &  6.32 & 0.105 & -0.59 & 8.69 & 1   \cr
NGC 6210    &  6.88 & 0.116 & -0.97 & 8.70 & 1   \cr
NGC 6309    &  5.61 & 0.118 & -0.84 & 8.91 & 1   \cr
NGC 6439    &  3.96 & 0.107 & -0.37 & 8.88 & 1   \cr
NGC 6543    &  7.69 & 0.112 & -0.78 & 8.72 & 1   \cr
NGC 6563    &  5.72 & 0.120 & -0.23 & 8.57 & 1   \cr
NGC 6565    &  6.11 & 0.101 & -0.44 & 8.91 & 1   \cr
NGC 6567    &  6.14 & 0.107 & -0.72 & 8.50 & 1   \cr
NGC 6572    &  6.97 & 0.119 & -0.57 & 8.77 & 1   \cr
NGC 6578    &  5.55 & 0.110 & -0.71 & 8.75 & 1   \cr
NGC 6629    &  6.03 & 0.100 & -0.88 & 8.61 & 1   \cr
NGC 6720    &  7.32 & 0.114 & -0.55 & 8.79 & 1   \cr
NGC 6790    &  6.49 & 0.105 & -0.62 & 8.60 & 1   \cr
NGC 6818    &  6.35 & 0.116 & -0.64 & 8.74 & 1   \cr
NGC 6826    &  7.55 & 0.103 & -0.93 & 8.43 & 1   \cr
NGC 6879    &  6.40 & 0.105 & -0.72 & 8.61 & 1   \cr
NGC 6884    &  7.56 & 0.117 & -0.70 & 8.66 & 1   \cr
NGC 6886    &  6.92 & 0.120 & -0.35 & 8.68 & 1   \cr
NGC 6891    &  6.57 & 0.113 & -0.97 & 8.65 & 1   \cr
NGC 6894    &  7.21 & 0.100 & -0.40 & 8.60 & 1   \cr
NGC 6905    &  6.93 & 0.104 & -0.58 & 8.66 & 1   \cr
NGC 7009    &  7.03 & 0.101 & -0.63 & 8.84 & 1   \cr
NGC 7026    &  7.64 & 0.113 & -0.45 & 8.79 & 1   \cr
NGC 7027    &  7.57 & 0.111 & -0.37 & 8.62 & 1   \cr
NGC 7354    &  7.88 & 0.095 & -0.43 & 8.92 & 5   \cr
NGC 7662    &  7.85 & 0.113 & -0.66 & 8.61 & 1   \cr
IC 351      & 10.36 & 0.100 & -0.95 & 8.48 & 1   \cr
IC 418      &  8.83 & 0.100 & -0.72 & 8.54 & 1   \cr
IC 1297     &  4.81 & 0.112 & -0.47 & 8.89 & 1   \cr
IC 1747     &  9.41 & 0.112 & -0.45 & 8.75 & 1   \cr
\ \cr
\noalign{\hrule}}}$$
\vfill\eject

\hskip 3.0 true cm Table II - Input data for PN (continued)
$$\vbox{\halign{#\hfil&\quad\hfil#\hfil&\quad\hfil#\hfil
&\quad\hfil#\hfil&\quad\hfil#\hfil&\quad\hfil#\hfil\cr\noalign{\hrule}
\ \cr
name & $R$ & He/H & $\log$(N/O) & $\epsilon$(O) & Ref. \cr
\ \cr
\noalign{\hrule}
\ \cr
IC 2003     &  9.83 & 0.094 & -0.58 & 8.62 & 1   \cr
IC 2149     &  8.65 & 0.099 & -0.24 & 8.54 & 1   \cr
IC 2165     &  9.08 & 0.105 & -0.37 & 8.42 & 1   \cr
IC 2448     &   7.35 & 0.095 & -0.31 & 8.59 & 1  \cr
IC 2501     &  7.49 & 0.093 & -0.77 & 8.92 & 1   \cr
IC 2621     &  7.10 & 0.090 & -0.42 & 8.90 & 1   \cr
IC 3568     &  8.68 & 0.099 & -0.87 & 8.57 & 1   \cr
IC 4406     &  6.44 & 0.116 & -0.85 & 8.75 & 2,4,7 \cr
IC 4776     &  4.39 & 0.090 & -1.08 & 8.86 & 1   \cr
IC 5117     &  7.66 & 0.110 & -0.57 & 8.61 & 1   \cr
IC 5217     &  8.56 & 0.100 & -0.68 & 8.70 & 1   \cr
Cn2-1       &  4.03 & 0.100 & -0.34 & 9.00 & 1   \cr
Fg 1        &  7.12 & 0.108 & -0.88 & 8.45 & 1   \cr
H1-32       &  4.92 & 0.110 & -1.00 & 8.58 & 8   \cr
H1-44       &  4.41 & 0.100 & -0.21 & 8.67 & 8   \cr
H1-56       &  4.01 & 0.100 & -1.10 & 8.72 & 8   \cr
H2-37       &  5.61 & 0.110 & -0.44 & 8.62 & 8   \cr
Hb 12       &  8.72 & 0.105 & -0.74 & 8.40 & 1   \cr
He2-21      &  9.96 & 0.116 & -0.86 & 8.45 & 2   \cr
He2-29      &  8.22 & 0.108 & -0.40 & 8.62 & 6   \cr
He2-37      &  7.76 & 0.119 & -0.59 & 8.96 & 1   \cr
He2-47      &  7.32 & 0.110 & -0.74 & 8.90 & 7,9 \cr
He2-48      &  8.15 & 0.106 & -0.59 & 8.53 & 1   \cr
He2-55      &  7.31 & 0.111 & -0.56 & 8.54 & 1   \cr
He2-67      &  7.14 & 0.120 & -0.87 & 8.91 & 7   \cr
He2-99      &  5.94 & 0.098 & -0.96 & 8.79 & 1   \cr
He2-115     &  6.17 & 0.100 & -0.87 & 8.62 & 1   \cr
He2-118     &  4.45 & 0.087 & -0.52 & 8.98 & 9   \cr
He2-123     &  5.53 & 0.117 & -0.11 & 8.67 & 1   \cr
He2-158     &  5.36 & 0.098 & -0.97 & 8.90 & 1   \cr
Hu1-1       & 10.70 & 0.105 & -0.60 & 8.68 & 1   \cr
J 320       & 11.46 & 0.107 & -0.87 & 8.33 & 1   \cr
J 900       &  9.65 & 0.098 & -0.90 & 8.60 & 1   \cr
K3-68       & 13.59 & 0.098 & -0.55 & 8.11 & 8   \cr
M1-1        & 10.65 & 0.114 & -0.60 & 8.30 & 1   \cr
M1-4        &  9.08 & 0.105 & -0.54 & 8.50 & 1   \cr
M1-5        &  9.69 & 0.100 & -1.00 & 8.54 & 1   \cr
M1-7        & 13.20 & 0.102 & -0.17 & 8.71 & 8   \cr
M1-14       & 10.27 & 0.099 & -0.98 & 8.40 & 3   \cr
M1-17       & 12.20 & 0.113 & -0.55 & 8.80 & 1,8 \cr
M1-25       &  4.04 & 0.117 & -0.67 & 8.99 & 1   \cr
M1-50       &  3.96 & 0.102 & -0.97 & 8.74 & 1   \cr
M1-54       &  4.62 & 0.112 & -0.19 & 8.97 & 1   \cr
M1-57       &  4.87 & 0.091 & -0.51 & 8.96 & 1   \cr
M1-60       &  4.31 & 0.105 & -0.26 & 8.84 & 9   \cr
\ \cr
\noalign{\hrule}}}$$
\vfill\eject

\hskip 3.0 true cm Table III - Input data for PN (continued)
$$\vbox{\halign{#\hfil&\quad\hfil#\hfil&\quad\hfil#\hfil
&\quad\hfil#\hfil&\quad\hfil#\hfil&\quad\hfil#\hfil\cr\noalign{\hrule}
\ \cr
name & $R$ & He/H & $\log$(N/O) & $\epsilon$(O) & Ref. \cr
\ \cr
\noalign{\hrule}
\ \cr
M1-74      &  6.38 & 0.099 & -0.60 & 8.78 & 1   \cr
M1-80      & 11.32 & 0.091 & -0.28 & 8.59 & 1   \cr
M2-2       &  8.81 & 0.100 & -0.48 & 8.43 & 1   \cr
M2-10      &  3.75 & 0.091 & -0.55 & 9.00 & 1   \cr
M2-27      &  5.31 & 0.120 & -0.43 & 8.89 & 8   \cr
M3-1       & 10.25 & 0.105 & -0.74 & 8.39 & 6   \cr
M3-4       & 10.62 & 0.123 & -0.51 & 8.72 & 6   \cr
M3-5       & 10.62 & 0.110 & -0.20 & 8.29 & 6   \cr
M3-6       &  8.68 & 0.090 & -1.27 & 8.64 & 1   \cr
M3-15      &  5.72 & 0.107 & -0.33 & 8.41 & 1   \cr
MaC 2-1    & 11.36 & 0.111 & -1.36 & 8.44 & 3   \cr
Pe1-18     &  6.31 & 0.091 & -0.37 & 8.92 & 1   \cr
Th2-A      &  6.44 & 0.092 & -0.60 & 8.74 & 1   \cr
\ \cr
\noalign{\hrule}}}$$

References: 

1 - Maciel and K\"oppen (1994)

2 - Costa {\it et al.} (1996) 

3 - Costa {\it et al.} (1997) 

4 - Corradi {\it et al.} (1997) 

5 - Hajian {\it et al.} (1997) 

6 - Kingsburgh and Barlow (1994)

7 - Perinotto {\it et al.} (1994) 

8 - Perinotto (1991)

9 - K\"oppen {\it et al.} (1991). 
\bigskip
\bigskip

The introduction of the correction procedure described in 
section~4 does not affect these results appreciably. As an example, 
Fig.~1 (bottom panel) shows the results for the van den Hoek 
and Groenewegen (1997) correction using the low mass calibration. 
Note that Fig.~1 gives the He/H abundances by number 
of atoms, as usual, so that the contamination $\Delta Y_s$ has to be
converted into the equivalent quantity by number $\Delta$(He/H).
The different calibrations lead to similarly negligible gradients,
which can be written as

$${d{\rm (He/H)} \over dR} = 0.0000 \pm 0.0004 \ . \eqno(6)$$

\noindent
In fact, the main effect of the corrections is to {\it reduce} 
the average He/H abundances: the van den Hoek and Gronewegen 
(1997) corrections lead to He/H $\simeq 0.091 \pm 0.003$ 
(high mass calibration) and He/H $\simeq 0.094 \pm 0.003$ (low mass 
calibration). Use of the Boothroyd and Sackmann (1999) 
data produces larger average  abundances by roughly 0.006, as the 
corrections are generally 
smaller. Since the He contamination is a function of the stellar mass, 
apparently central stars with different masses are scattered 
homogeneously in the whole range of galactocentric distances, thus 
destroying any systematic variations. It can be concluded that, 
{\it in view of the uncertainties involved both in the abundances 
and distances, it is unlikely that any He/H radial gradient could 
be presently  detected from planetary nebulae.} 
The fact that some previous determinations led to a small gradient 
is probably due to the use of small samples, as in the case of the 
39 PN by Fa\'undez-Abans and Maciel (1986). Determinations of the 
He/H gradient based on HII regions may also be affected by small 
samples. In this case, the distances are usually better known and 
no corrections are needed for stellar contamination, but the
abundance determinations may depend on somewhat uncertain corrections 
for the presence of neutral helium in the nebulae (cf. Peimbert 
1979).

The present results can be used to estimate an {\it upper limit} 
to the He/H gradient on the basis of the total dispersion 
$\sigma_d$ observed. Since $\sigma_d \simeq 0.04$ in all cases 
considered, we have

$$\bigg\vert {d({\rm He/H}) \over dR}\bigg\vert  < {\sigma_d \over \Delta R}
    \simeq  0.004 \ {\rm kpc}^{-1} \eqno(7)$$

\noindent
corresponding to $d\log({\rm He/H})/dR \simeq -0.02$ dex/kpc
for $\Delta R \simeq 10$~kpc. {\it Therefore, the existence of a
He/H radial gradient with magnitude similar to the O/H 
gradient is extremely unlikely, so that any He/H gradient should
be lower than the O/H gradient by at least a factor of 3.} This 
conclusion is in good agreement with the small gradients recently 
derived for galactic HII regions by Esteban {\it et al.} 
(1999), as discussed in section~2.

Recent chemical evolution models support the present conclusions. 
Models by Matteucci and co-workers (Matteucci and Fran\c cois 
1989, Matteucci and Chiappini 1999) predict 
very small gradients given by $d\log({\rm He/H})/dR \simeq -0.0085$ dex/kpc, 
approximately $d({\rm He/H})/dR \simeq -0.002$ kpc$^{-1}$. The recent 
models by Chiappini {\it et al.} (1997) imply a maximum 
gradient for the inner Galaxy of $d\log({\rm He/H})/dR = -0.006$ dex/kpc, 
with values smaller by a factor 3 for the outer region, so that these 
results are consistent with the present investigation.
\bigskip
\bigskip\noindent 
{5.2 THE He TO METALS ENRICHMENT RATIO}
\medskip\noindent
We have also applied the procedure of section~4 to the PN sample 
of Maciel and Quireza (1999), taking the O/H abundance as 
representative of the total metallicity. We adopted the relation 
$Z \simeq 25 \ $O/H, where O/H is the oxygen abundance by number 
relative do hydrogen, so that oxygen comprises 45\% of the total 
abundances by mass.  As shown by Chiappini and Maciel (1994), 
this is essentially the same as their independently derived relation for 
type~II PN. Since it can also be applied to HII regions (see for 
example Peimbert 1990), the conversion from O/H to Z for all
photoionized nebulae becomes straightforward.

The simple linear relationship~1 is certainly valid for 
metal-poor objects such as blue compact galaxies and low metallicity 
HII regions, but it probably breaks up at some maximum metallicity, 
after which a more sophisticated model would be necessary. As pointed 
out by Chiappini and Maciel (1994), He abundances of PN show 
some tendency to flatten out for $Z \geq 0.010$, which was taken in
their work as an upper limit for the application of Eq.~1. 
Moreover, some large metallicity PN have larger than average N/O ratios, 
apparently due to ON cycling in the progenitor stars. For these objects 
the necessary corrections to the He abundance are also larger, and 
therefore more uncertain. In this work, we decided to extend a
little further the limit set by Chiappini and Maciel (1994)
and take into account all PN in our sample having metal abundances
up to $10^6\, {\rm O/H} \simeq 700$, which corresponds approximately
to the solar value, $\epsilon({\rm O}) = \log {\rm (O/H)} + 12
= 8.83$ (Grevesse and Sauval 1998), or $Z \simeq 0.017$ 
if we take $Z = 25 \, $O/H. This includes 81 objects, or about 80\% of
the PN in our sample, so that it is still a representative
sample of the planetary nebulae in the galactic disk.

Our main goal here is to determine the helium to heavy
element enrichment ratio $\Delta Y/\Delta Z$. The pregalactic
He abundance $Y_p$ could also be obtained using the PN alone,
as in some previous work (see for example Maciel 1988), 
but this quantity is clearly better determined on the basis of very low 
metallicity objects, as mentioned in section~3. Therefore, we 
have taken $Y_p$ as a parameter, with the values $Y_p = 0.23$ 
and $Y_p = 0.24$. For the purposes of the present investigation, 
the discussion on the third decimal place is largely irrelevant. 

We have then applied Eq.~1 to the PN sample and obtained plots
of the He abundance by mass $Y$ both as a function of the O/H abundance
by number and the total metal abundance by mass $Z$, as shown in
Figs.~2 and 3 (filled circles). Fig.~2 
gives the uncorrected abundances and fits, and Fig.~3 
shows the results using the corrections according to the van den Hoek 
and Groenewegen data. In both figures, the straight lines are least 
squares fits using $Y_p = 0.23$ (dashed lines)  and $Y_p = 0.24$ 
(solid lines).
\bigskip
%
\centerline{\psfig{figure=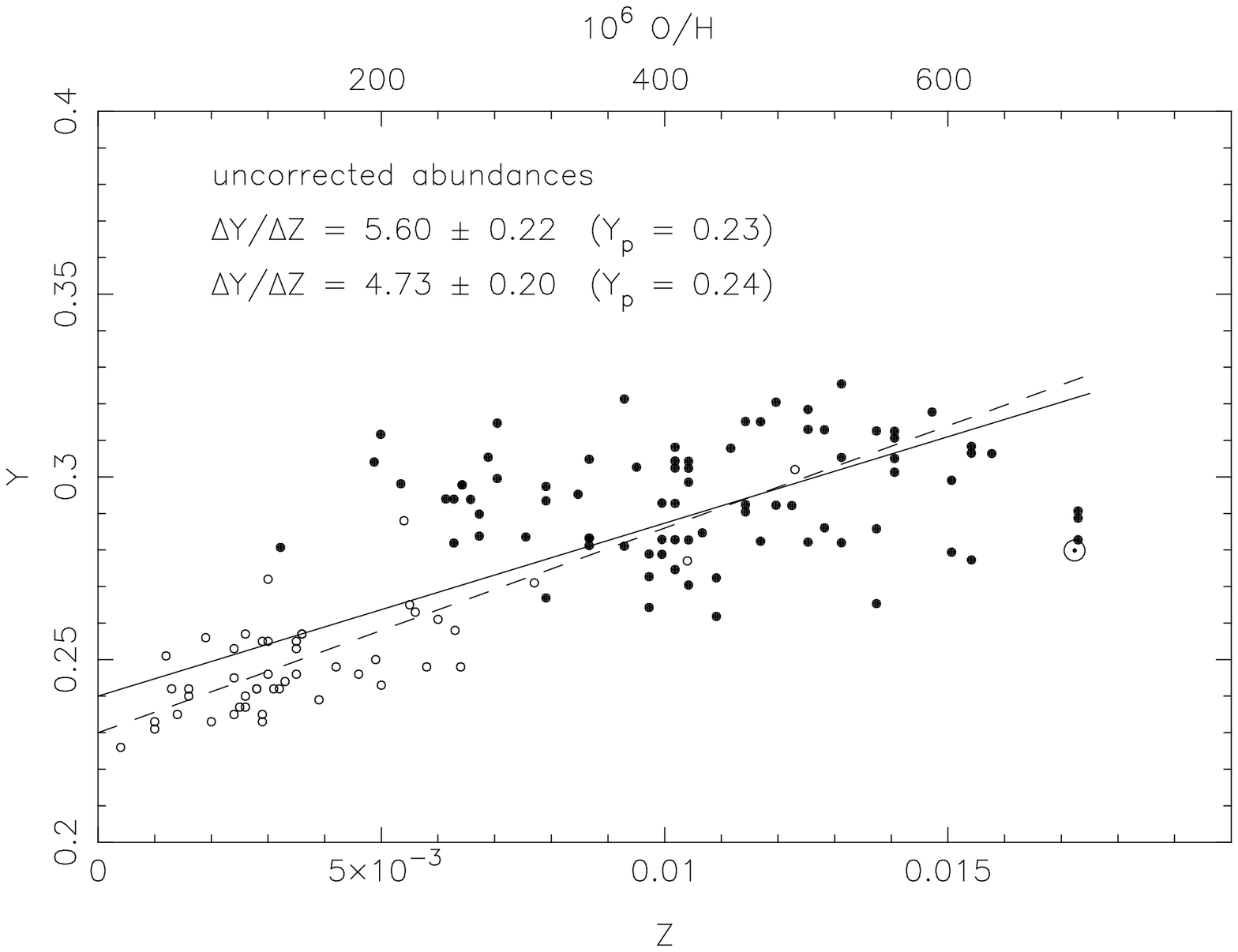,height=8.0 cm,angle=0} }
\noindent{\sevenrm \baselineskip 12 pt Figure 2 - 
             Uncorrected He abundances by mass 
             $Y$ for PN (solid 
             circles) as a function of the O/H abundances by number 
             (top axis) and total heavy element abundance by mass $Z$ 
             (bottom axis). Also shown are the Sun ($\odot$) and HII 
             regions (empty circles). The straight lines are least 
             squares fits for $Y_p = 0.23$ (dashed line) and $Y_p = 0.24$ 
             (solid line).}
%
\bigskip
\bigskip

In order to have a better idea of the behaviour of the $Y(Z)$
function at low and high metallicities, Figs.~2 and 
3 also include the Sun ($\odot$) and a sample of HII regions 
and metal poor blue compact galaxies (empty circles) from the 
compilation of Chiappini and Maciel (1994). These objects 
have {\it not} been taken into account in the determination of the 
linear fits, and are included for comparison purposes only. In fact, 
inclusion of these low metallicity objects would not affect the derived 
slopes by more than a few percent, since the main constraint at low 
metallicities is defined by the pregalactic abundance $Y_p$. 

From Figs.~2 and 3 it can be seen that the scatter 
of the low metallicity objects is lower than for the PN, which is 
confirmed by considering different samples of metal poor objects in the
literature. The higher scatter in the PN data is partially due to the 
higher uncertainties in the abundances and also to the correction 
procedure adopted in section~4, both of which may be improved in the 
future. In fact, the determination of abundances will benefit from
the use of more realistic photoionization models and better empirical
formulae including improved icf's. The correction procedure can 
be also improved as better stellar models are available, which
will allow a more accurate determination of the excess helium
as a function of the observed abundances and stellar mass. 
The higher uncertainties in the PN abundances and the fact that 
the He/H gradient is at most a factor of 3 lower than the O/H gradient 
are probably responsible for the lack of a He/H $\times$ $R$ correlation, 
while some  correlation between $Y$ and $Z$ (or O/H) can be observed, 
particularly from Fig.~3. In fact, the application of the
correction procedure not only reduces the average He abundances and
the derived $\Delta Y/\Delta Z$ ratio, but also decreases the 
uncertainty of the derived slopes, as can be seen by a comparison of
Figs.~2 and 3. This effect is similar for both
calibrations, being slightly stronger for the high mass calibration,
as can be seen in the top panel of Fig.~3.
\bigskip\bigskip
%
\centerline{\psfig{figure=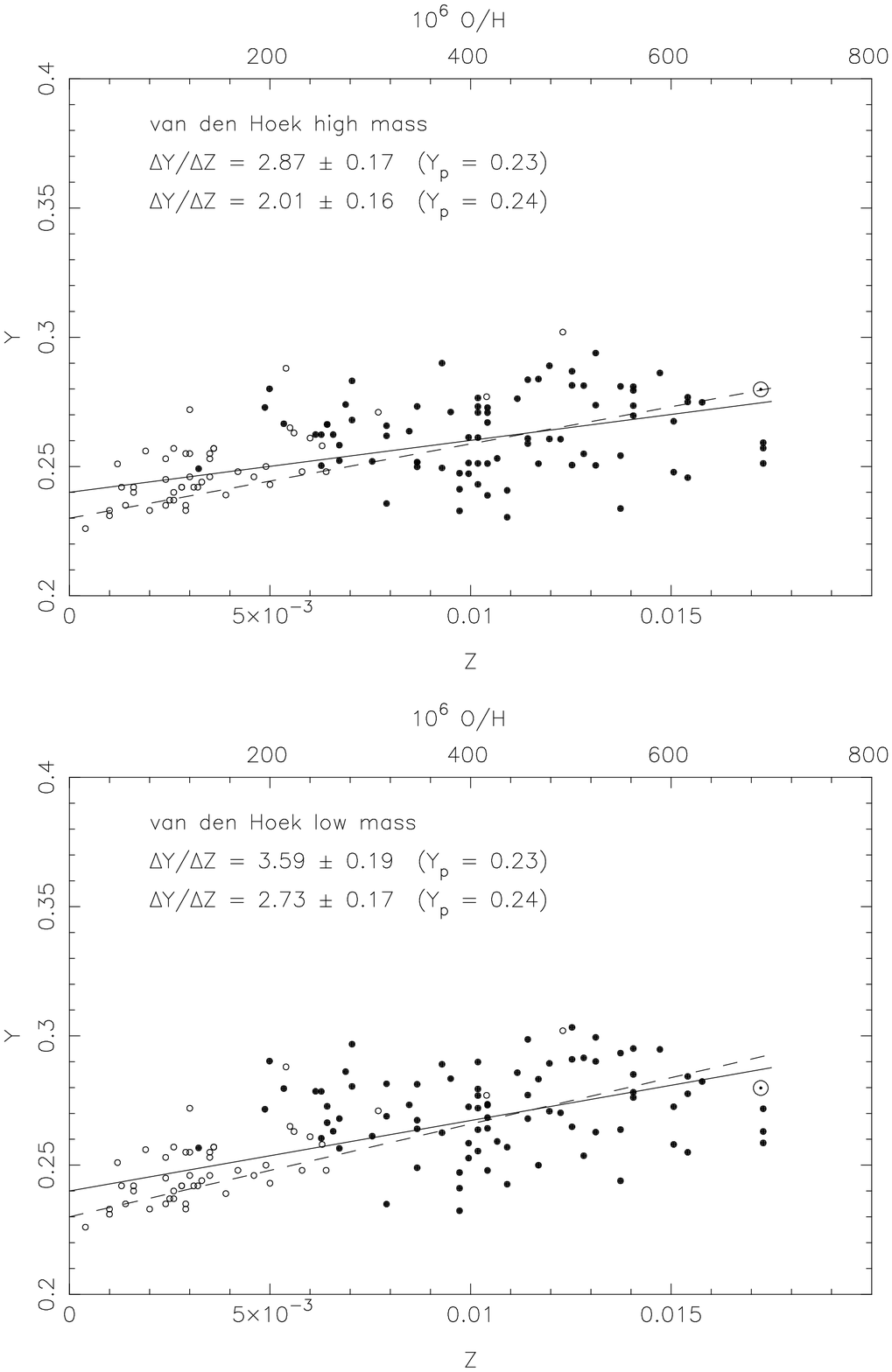,height=16.0 cm,angle=0} }
\noindent{\sevenrm \baselineskip 12 pt 
            Figure 3 - The same as Fig.~2 using corrected abundances 
            according to the van den Hoek and Groenewegen (1997) 
            data for the high mass calibration (top panel) and low mass 
            calibration (bottom panel).}
%
\bigskip

The detailed results of the different calibrations are shown in 
Table~IV.  For easier comparison with other investigations, this 
table also gives the slopes $\Delta Y/\Delta({\rm O/H})$ and 
$\Delta Y/\Delta Z_{16}$ relative to the oxygen abundance by number 
and by mass, respectively, which are given by

$${\Delta Y \over \Delta Z_{16}} = {1 \over 0.45}\
    {\Delta Y \over \Delta Z} = {1 \over 25 \times 0.45}\
    {\Delta Y \over \Delta ({\rm O/H})} \ . \eqno(8)$$

\medskip\noindent
Note that our ratio $\Delta Y/\Delta Z_{16}$ is the same
as the ratio $\Delta Y/\Delta {\rm O}$ as defined by
Peimbert (1995). As discussed by Esteban et al. 
(1999), the ratio  $\Delta Y/ \Delta Z_{16}$ is a better 
constraint to chemical evolution models, since no corrections for 
the remaining elements are needed. They derive ratios in the range 
$\Delta Y/ \Delta Z_{16} = 4.78$ to 6.55, with an average 
$\Delta Y/ \Delta Z_{16} = 5.87$ for $Y_p = 0.240 \pm 0.005$. From 
Table~IV we have $\Delta Y/ \Delta Z_{16} \simeq 5.27$ 
for $Y_p = 0.24$, so that the agreement is very good.
\bigskip

\hskip 3.0 true cm Table IV - Results for the $\Delta Y/\Delta Z$ ratio
$$\vbox{\halign{#\hfil&\quad\hfil#\hfil&\quad\hfil#\hfil
&\quad\hfil#\hfil&\quad\hfil#\hfil&\quad\hfil#\hfil
&\quad\hfil#\hfil\cr\noalign{\hrule}
\ \cr
  & uncorrected  &  high $M$ & low $M$  \cr
\ \cr
\noalign{\hrule}
\ \cr
$Y_p = 0.23$    &       &       &                \cr
$\Delta Y/
\Delta$(O/H)    & $139.9\pm 5.6$ & $71.9\pm 4.2$ & $89.7\pm 4.7$  \cr
$\Delta Y/\Delta Z_{16}$  & $12.44\pm 0.50$  &  $6.39\pm 0.37$ 
   &  $7.98\pm 0.41$   \cr
$\Delta Y/
\Delta Z$       & $5.60\pm 0.22$  &  $2.87\pm 0.17$ &  $3.59\pm 0.19$  
     \cr
corr. coeff.        & $0.94$  & $0.89$  & $0.91$  \cr
                &       &       &              \cr
$Y_p = 0.24$    &       &       &                \cr
$\Delta Y/
\Delta$(O/H)    & $118.3\pm 5.1$ & $50.2\pm 3.9$ & $68.1\pm 4.3$  
     \cr
$\Delta Y/\Delta Z_{16}$  & $10.52\pm 0.45$  &  $4.47\pm 0.35$ 
   &  $6.06\pm 0.38$  \cr
$\Delta Y/
\Delta Z$       & $4.73\pm 0.20$  &  $2.01\pm 0.16$ &  $2.73\pm 0.17$  
    \cr
corr. coeff.        & $0.93$  & $0.82$  & $0.87$ \cr
\ \cr
\noalign{\hrule}}}$$
\bigskip
\bigskip

The $\Delta Y/\Delta Z$ ratio decreases from the range  
$4.7 < \Delta Y/\Delta Z < 5.6$, which is similar to the result 
of Chiappini and Maciel (1994), to the values

$$2.8 < {\Delta Y\over \Delta Z} < 3.6 \qquad (Y_p = 0.23) \eqno(9)$$

$$2.0 < {\Delta Y\over \Delta Z} < 2.8 \qquad (Y_p = 0.24)\ . \eqno(10)$$

\noindent
These  results are closer to independently derived ratios, as
we have seen in section~3. If we consider the Boothroyd and Sackmann
data, these values are increased by 35\% approximately.

Predicted values for the enrichment ratio are usually lower
than $\Delta Y/\Delta Z \simeq 3$, a condition that can be
fulfilled by the present results both for $Y_p = 0.23$ and 
$Y_p = 0.24$. Recent models by Allen {\it et al.} (1998) 
and also models based on two main infall episodes for the formation of
the thin disk and halo/thick disk (Chiappini {\it et al.} 
1997) predict ratios as low as $\Delta Y/\Delta Z = 1.6$, 
which is closer to our lower limit for $Y_p = 0.24$.

It should also be noted that, while the derived $\Delta Y/\Delta Z$ 
ratios are similar to recent determinations in the literature, the 
average {\it uncertainties} are smaller, which is a consequence 
of the fact that a large number of objects has been included, with
a larger metallicity spread, leading to a more reliable 
determination. As an example, the recent value by Pagel and Portinari 
(1998) is $\Delta Y/\Delta Z = 3 \pm 2$, and results from 
blue compact dwarf galaxies by Thuan and Izotov (1998) are 
$\Delta Y/\Delta Z = 2.3 \pm 1.0$. An average including  both 
calibrations of Table~IV would give $\Delta Y/\Delta Z = 
3.2 \pm 0.5$ for $Y_p = 0.23$ and  $\Delta Y/\Delta Z = 2.4 \pm 0.5$ 
for $Y_p = 0.24$. These results are similar to the recent determinations 
of Esteban {\it et al.} (1999) for the HII regions M8, 
M17 and Orion assuming temperature fluctuations characterized by a $t^2$ 
parameter greater than zero. It may be remarked that 
both the uncorrected results of Table~IV and the 
ratios derived by Esteban {\it et al.} (1999) for 
$t^2 = 0.0$ are larger by roughly 50\%. According to Peimbert 
(1995), temperature fluctuations and the possibility 
of grain condensation imply some increase in the oxygen abundance for 
a given value of the He abundance, decreasing the slope 
$\Delta Y/\Delta Z$. The same effect is attained here by correcting 
the He abundance by a certain amount $\Delta Y_s$.
Tables~I  to III
include several objects for which relatively large temperature
fluctuation parameters $t^2$ are estimated, such as NGC~2392,
NGC~3211 and NGC~3242 (Liu and Danziger 1993). These
objects are a small fraction of the sample, and display the
same dispersion in Figs. 1 to 3 as the remaining objects,
confirming that the adopted correction procedure is also adequate
for these nebulae. 

It is interesting to note from Fig.~3 that there seems to be
some continuity from the very low metallicity blue compact galaxies
($10^6 $ O/H $< 120$) to the HII regions ($10^6 $ O/H $< 300$)
and disk PN ($10^6 $ O/H $> 120$). This might seem surprising,
as these objects have apparently had different chemical evolution
histories. However, it could be argued that the chemical evolution
of a system is basically defined by its initial mass function (IMF)
and star formation history (SFH). The IMF is now believed to be
universal, or at least to have only small local variations
(see for example Padoan {\it et al.} 1997, Maciel and 
Rocha-Pinto 1998). The SFH is clearly \lq\lq bursty\rq\rq\ 
in  blue compact galaxies, but it is generally assumed as constant in the 
Galaxy. However, recent work by Rocha-Pinto {\it et al.} 
(2000) has shown that our Galaxy presents clear evidences 
of past enhanced star formation periods, or bursts, followed by periods 
of depressed star formation, or lulls. As a consequence, only the 
{\it average} star formation rate can be considered as constant, so 
that the similarity in the chemical evolution of the different systems 
shown in Fig.~3 is probably not surprising at all.

\bigskip
\centerline{\bf Acknowledgements}
\medskip\noindent
This work was partially supported by CNPq and FAPESP.
\bigskip
\centerline{\bf REFERENCES}
\medskip
\ref
Allen, C., Carigi, L. and Peimbert, M.: 1998, 
    {\it Astrophys. J. \bf 494}, 247.
\ref
Amnuel, P.R.: 1993, {\it Monthly Notices Roy. Astron.
     Soc. \bf 261}, 263.
\ref
Bl\"ocker, T. and Sch\"onberner, D.: 1990, 
    {\it Astron. Astrophys. \bf 240}, L11.
\ref
Boothroyd, A.I. and Sackmann, I.-J.: 1988, 
    {\it Astrophys. J. \bf 328}, 653.
\ref
Boothroyd, A.I. and Sackmann, I.-J.: 1999, 
    {\it Astrophys. J. \bf 510}, 232.
\ref
Burles, S., Nollett, K.M., Truran, J.N. 
    and Turner, M.S.: 1999, {\it Phys. Rev. Lett. \bf 82}, 4176.
\ref
Cazetta, J.O. and Maciel, W.J.: 2000, {\it Rev. 
    Mexicana Astron. Astrof. \bf 36}, 3
\ref
Chiappini, C. and Maciel, W.J.: 1994, 
    {\it Astron. Astrophys. \bf 288}, 921.
\ref
Chiappini, C., Matteucci, F. and Gratton, R.: 
    1997, {\it Astrophys. J. \bf 477}, 765.
\ref
Corradi, R.L.M., Perinotto, M., Schwarz, H.E. 
    and Claeskens, J.F.: 1997, {\it Astron. Astrophys. \bf 322}, 975.
\ref
Costa, R.D.D., Chiappini, C., Maciel, W. J. 
    and Freitas-Pacheco, J.A.: 1996, {\it Astron. Astrophys. Suppl. \bf 116}, 
    249.
\ref
Costa, R.D.D., Chiappini, C., Maciel, W. J. 
    and Freitas-Pacheco, J.A.: 1997, {\it Advances in stellar evolution}, 
    ed. R.T. Rood, A. Renzini, Cambridge, 159.
\ref
D'Odorico, S., Peimbert, M. and Sabbadin, F.: 
    1976, {\it Astron. Astrophys. \bf 47}, 341.
\ref
Efremov, Y.N., Schilbach, E. and Zinnecker, H.: 
    1997, {\it Astron. Nachr. \bf 6}, 335.
\ref
Esteban, C. and Peimbert, M.: 1995, {\it Rev. 
    Mexicana Astron. Astrof. SC \bf 3}, 133.
\ref
Esteban, C., Peimbert, M., Torres-Peimbert, 
    S. and Garcia-Rojas, J.: 1999, {\it Rev. Mexicana Astron. Astrof. 
    \bf 35}, 65.
\ref
Fa\'undez-Abans and M., Maciel, W.J.: 1986, {\it
    Astron. Astrophys. \bf 158}, 228.
\ref
Grevesse, N. and Sauval, A.J.: 1998, {\it Space Sci. 
    Rev. \bf 85}, 161. 
\ref
Groenewegen, M.A.T. and de Jong, T.: 1993, {\it
    Astron. Astrophys. \bf 267}, 410.
\ref
Groenewegen, M.A.T., van den Hoek, L.B. and de 
    Jong, T.: 1995, {\it Astron. Astrophys. \bf 293}, 381.
\ref
Gruenwald, R.B. and Viegas, S.M.M.: 1998, 
    {\it Astrophys. J. \bf 501}, 221.
\ref
Hajian, A.R., Balick, B., Terzian, Y. and 
    Perinotto, M.: 1997, {\it Astrophys. J. \bf 487}, 313.
\ref
H\o g, E., Pagel, B.E.J., Portinari, L., 
    Thejll, P.A., Macdonald, J. and Girardi, L.: 1998, {\it Space Sci. 
    Rev. \bf 84}, 115.
\ref
Izotov, Y.I., Thuan, T.X. and Lipovetsky, V.A.: 
    1997, {\it Astrophys. J. Suppl. \bf 108}, 1.
\ref
Izotov, Y.I., Chaffee, F.H., Foltz, C.B., 
    Green, R.F., Guseva, N.G. and Thuan, T.X.: 1999, {\it Astrophys. J. 
    \bf 527}, 757.
\ref
Kingsburgh, R.L. and Barlow, M.J.: 1994, 
    {\it Monthly Notices Roy. Astron. Soc. \bf 271}, 257.
\ref
K\"oppen, J., Acker, A. and Stenholm, B.: 1991, 
    {\it Astron. Astrophys. \bf 248}, 197.
\ref
Lequeux, J., Peimbert, M., Rayo, J.F., 
    Serrano, A. and Torres-Peimbert, S.: 1979, {\it Astron. Astrophys. 
    \bf 80}, 155.
\ref
Liu, X.W. and Danziger, J.: 1993, {\it Monthly 
    Notices Roy. Astron. Soc. \bf 263}, 256.
\ref
Maciel, W.J.: 1988, {\it Astron. Astrophys. 
    \bf 200}, 178.
\ref
Maciel, W.J.: 1993, {\it Astrophys. Space.
    Sci. \bf 206}, 285.
 \ref
Maciel, W.J.: 1997, {\it IAU Symp. 180}, ed. 
    H.J. Habing, H.J.G.L.M. Lamers, Kluwer, 397.
 \ref
Maciel, W.J.: 2000, {\it Chemical evolution of 
    the Milky Way: stars versus clusters}, ed. F. Giovannelli, F. 
    Matteucci, Kluwer (in press)
\ref
Maciel, W.J. and Cazetta, J.O.: 1997, {\it
    Astrophys. Space Sci. \bf 249}, 341.
\ref
Maciel, W.J. and Chiappini, C.: 1994, 
    {\it Astrophys. Space Sci. \bf 219}, 231.
\ref
Maciel, W.J. and K\"oppen, J.: 1994, {\it
    Astron. Astrophys. \bf 282}, 436.
 \ref
Maciel, W.J. and Quireza, C.: 1999, {\it
    Astron. Astrophys. \bf 345}, 629.
\ref
Maciel, W.J. and Rocha-Pinto, H.J.: 1998, {\it
    Monthly Notices Roy. Astron. Soc. \bf 299}, 889.
\ref
Maeder, A.: 1992, {\it Astron. Astrophys. 
    \bf 264}, 105.
\ref
Marigo, P., Bressan, A. and Chiosi, C.: 1996, 
    {\it Astron. Astrophys. \bf 313}, 545.
\ref
Marigo, P.: 2000, {\it Chemical evolution of 
    the Milky Way: stars versus clusters}, ed. F. Giovannelli, F. 
    Matteucci, Kluwer (in press)
 \ref
Matteucci, F. and Chiappini, C.: 1999, 
    {\it Chemical evolution from zero to high redshift}, ed. J.R. Walsh, 
    M.R. Rosa, Springer, 83
 \ref
Matteucci, F. and Fran\c cois, P.: 1989, 
    {\it Monthly Notices Roy. Astron. Soc. \bf 239}, 885.
\ref
M\'endez, R.H., Kudritzki, R.P., Herrero, 
    A., Husfeld, D. and Groth, H.G.: 1988, {\it Astron. Astrophys. \bf 
    190}, 113.
\ref
M\'endez, R.H., Kudritzki, R.P. and Herrero, 
    A.: 1992, {\it Astron. Astrophys. \bf 260}, 329.
\ref
Olive, K.A., Steigman, G. and Walker, T.P.: 
    2000, {\it Phys. Rep.} (in press)
\ref
Padoan, P., Nordlund, \AA. and Jones, B.J.T.: 1997, 
    {\it Monthly Notices Roy. Astron. Soc. \bf 288}, 145.
\ref
Pagel, B.E.J.: 1989, {\it Evolutionary phenomena 
    in galaxies}, ed. J.E. Beckman, B.E.J. Pagel, Cambridge, 368. 
\ref
Pagel, B.E.J.: 1995, {\it The light element 
    abundances}, ed. P. Crane, Springer, 155.
\ref
Pagel, B.E.J. and Portinari, L.: 1998, 
    {\it Monthly Notices Roy. Astron. Soc. \bf 298}, 747.
\ref
Pagel, B.E.J., Simonson, E.A., Terlevich, 
    R.J. and Edmunds, M.C.: 1992, {\it Monthly Notices Roy. Astron. Soc.
    \bf 255}, 325.
\ref
Pasquali, A. and Perinotto, M.: 1993, 
    {\it Astron. Astrophys. \bf 280}, 581.
\ref
Peimbert, M.: 1978, {\it IAU Symp. 76}, ed. 
    Y. Terzian, Reidel, 215.
\ref
Peimbert, M.: 1979, {\it IAU Symp. 84}, ed. 
    W.B. Burton, Reidel, 307.
\ref
Peimbert, M.: 1990, {\it Rep. Prog. Phys.
    \bf 53}, 1559.
\ref
Peimbert, M.: 1995, {\it The light element 
    abundances}, ed. P. Crane, Springer, 165.
\ref
Peimbert, M. and Serrano, A.: 1980, 
    {\it Rev. Mexicana Astron. Astrof. \bf 5}, 9.
\ref
Peimbert, M. and Torres-Peimbert, S.: 1974, 
    {\it Astrophys. J. \bf 193}, 327.
\ref
Peimbert, M. and Torres-Peimbert, S.: 1976, 
    {\it Astrophys. J. \bf 203}, 581.
\ref
Perinotto, M.: 1991, {\it Astrophys. J. 
    Suppl. \bf 76}, 687.
\ref
Perinotto, M., Purgathofer, A., 
    Pasquali, A. and Patriarchi, P.: 1994,  {\it Astron. Astrophys. Suppl.
    \bf 107}, 495.
\ref
Renzini, A. and Voli, M.: 1981, 
    {\it Astron. Astrophys. \bf 94}, 175.
\ref
Rocha-Pinto, H.J., Scalo, J., Maciel, W.J. 
    and Flynn, C.: 2000, {\it Astrophys. J. Lett. \bf 531}, L115. 
\ref
Sackmann, I.-J. and Boothroyd, A.I.: 1999, 
    {\it Astrophys. J. \bf 510}, 217.
\ref
Sch\"onberner, D.: 1983, {\it Astrophys. J.
    \bf 272}, 708.
\ref
Shaver, P.A., McGee, R.X., Newton, L.M., 
    Danks, A.C. and Pottasch, S.R.: 1983, {\it Monthly Notices Roy.
    Astron. Soc. \bf 204}, 53.
\ref
Stasi\' nska, G. and Tylenda, R.: 1990, 
    {\it Astron. Astrophys. \bf 240}, 467.
\ref
Stasi\' nska, G., Gorny, S.K. and
    Tylenda, R.: 1997, {\it Astron. Astrophys. \bf 327}, 736.
\ref
Thuan, T.X. and Izotov, Y.I.: 1998, {\it Space 
    Sci. Rev. \bf 84}, 83.
\ref
van den Hoek, L.B. and Groenewegen, M.A.T.: 
    1997, {\it Astron. Astrophys. Suppl. \bf 123}, 305.
\ref
Weidemann, V. and Koester, D.: 1983, 
    {\it Astron. Astrophys. \bf 121}, 77.
\ref
Zhang, C.Y.: 1993, {\it Astrophys. J. \bf 410}, 239.
\ref
\bye